\documentclass[12pt]{iopart}
\newcommand{\forget}[1]{}

\usepackage{graphicx}

\begin{document}

\title[FD in sub-diffusive systems]{Fluctuation-Dissipation relation in sub-diffusive systems: the case of granular single-file}

\author{D. Villamaina}
\address{Universit\`a di Roma ``La Sapienza'', Dipartimento di
Fisica, p.le Aldo Moro 2, I-00185 Roma, Italy}

\author{A. Puglisi}
\address{CNISM and Universit\`a di Roma ``La Sapienza'', Dipartimento di
Fisica, p.le Aldo Moro 2, I-00185 Roma, Italy}

\author{A. Vulpiani}
\address{Universit\`a di Roma ``La Sapienza'', Dipartimento di
Fisica, CNISM and INFN, p.le Aldo Moro 2, I-00185 Roma, Italy}

\ead{andrea.puglisi@roma1.infn.it}

\begin{abstract}
  We study a gas of hard rods on a ring, driven by an external thermostat, with
  either elastic or inelastic collisions, which exhibits sub-diffusive
  behavior $\langle x^2 \rangle \sim t^{1/2}$. We show the validity of
  the usual Fluctuation-Dissipation (FD) relation, i.e. the
  proportionality between the response function and the correlation
  function, when the gas is elastic or diluted. On the contrary, in
  strongly inelastic or dense cases, when the tracer velocity is no
  more independent of the other degrees of freedom, the Einstein
  formula fails and must be replaced by a more general FD relation.
\end{abstract}


\maketitle

\paragraph{Introduction}

The typical scenario in diffusive problem is the so called standard
diffusion, which is qualitatively similar
to the usual behaviour in the Brownian motion,
 i.e. at large time one has
\begin{equation}
\label{1}
\langle x(t)^2 \rangle \simeq 2D t \,\, ,
\end{equation}
where $D$ is the diffusion coefficient which is linked to the
velocity correlation function via the Kubo formula
$$
D=\int_0^{\infty} \langle v(t)v(0) \rangle dt \, .
$$
Of course the above scenario holds if $\int_0^{\infty} \langle v(t)v(0) \rangle dt$
is finite and non-zero.

On the other hand, it is  well known that, beyond the
standard diffusion, one can have anomalous diffusion~\cite{ZSW93,CMMV99},  i.e.
\begin{equation}
\label{2}
\langle x^2(t)\rangle \sim t^{2 \nu} \,\,\, \mbox{with} \,\,\,  \nu \ne
1/2,
\end{equation}
formally this corresponds to have $D=\infty$ if $\nu > 1/2$
(superdiffusion)
and $D=0$ if $\nu < 1/2$ (subdiffusion).

From the  well estabished linear response theory,
it is known that, when $\langle x(t)\rangle=0$ in the unperturbed system,
 (\ref{1}) implies  a linear drift
\begin{equation}
\label{3}
\overline{x(t)} \sim t \,\,,
\end{equation}
if a small external force is applied~\cite{K66,BPRV}. In the following we
will indicate with $\langle \cdot \rangle$ the average in the
unperturbed system, i.e. weighting states according to the stationary 
phase-space distribution and with $\overline{(\cdot)}$ the time dependent average in the
dynamical ensemble generated by the external perturbation. One can
wonder how Eq.~(\ref{3}) changes in presence of anomalous diffusion,
i.e. if, instead of (\ref{1}), eq. (\ref{2}) holds. 

The ``usual'' fluctuation-dissipation relation relates the mean
response $R(t)=\frac{\overline{\delta v(t)}}{\delta v(0)}$ at time $t$
of the velocity after an impulsive infinitesimal perturbation $\delta
v(0)$, applied at time $t=0$, to the velocity autocorrelation
$C_v(t_1-t_2)=\langle v(t_1)v(t_2) \rangle$:
$$
R(t)= C_v(t)/C_v(0).
$$
When an infinitesimal force is applied for positive times,
one has 
\begin{equation}
\label{A.5}
\overline{v(t)}={d \over dt} \overline{x(t)} \propto \int_0^t
 C_v(t')  dt' \,\,.
\end{equation}
A straigthforward consequence of the above relations and of the simple identity 
\begin{equation}
\label{A.4}
\langle x^2(t) \rangle =\int_0^t\int_0^t
C_v(t_1-t_2)dt_1 dt_2  \,\,\,\, , \,\,\,\,
\end{equation}
which suggests 
\begin{equation}
\label{A.7}
\overline{x(t)}=\int_0^t \overline{v(t)} dt \propto \langle x^2(t) \rangle \sim t^{2\nu} \,\, ,
\end{equation}
in analogy with~(\ref{3}).
On the other hand, it can be seen that such a formal
argument is not rigorous  and the actual scenario may
become rather subtle, see e.g.~\cite{TFWG94}.  For a detailed
discussion the reader can see~\cite{BPRV}.

In this paper we discuss the subdiffusive situation.
Some works  show that in such a case 
the expected result (\ref{A.7}) seems to hold~\cite{BK98,MBK99}.
This has been explicitely proved in systems described by a
fractional-Fokker-Planck~\cite{MBK99} equation, where a 
generalized Einstein relation has been shown ($F$ is the perturbing force).
\begin{equation}
\label{A.16}
\overline{x(t)}={1 \over 2} {F\langle x^2(t) \rangle \over k_B T} \,\, .
\end{equation}

Models based on fractional Fokker-Planck equations, although
interesting, usually are not directly derived from specific real
systems; we therefore wondered whether a relation similar
to~(\ref{A.16}) holds in more realistic models, such as in single file
diffusion~\cite{L73}, which is a sub-diffusive system having many
realizations in nature (e.g. transport in nanopores or narrow
channels and zeolites, as well as car traffic on single lanes, pedestrian
dynamics, etc.). The model used here consists of a one-dimensional
gas of inelastic hard particles, moving on a large ring. To ensure a
stationary state, particles exchange energy with an external
thermostat.  Tuning the characteristic time of the thermostat, the
average volume fraction occupied by the gas and the restitution
coefficient (from elastic to completely anelastic), one may observe a
wide range of different stationary states, from a homogeneous density
with Gaussian velocity distribution to strongly inhomogeneous spatial
arrangement (clustering) with non-Gaussian statistics of
velocities~\cite{WM96,PLMPV98,PLMV99,NE98,NETP99}.  Other authors
have studied diffusion in granular gases without any external driving:
in this case the gas is non-stationary (cooling regime) and one finds
non-trivial exponents for diffusion\forget{, for instance $\nu=1/12$
in the homogeneous cooling regime for viscoelastic particles and even
a logarithmic diffusion for a constant inelasticity}~\cite{BP00b,BRCG00}.

The aim of this paper is to discuss the consequences of both
subdiffusion and inelasticity in the more general context of linear
response theory for statistically stationary states~\cite{DH75,FIV90}.
Let us briefly remind some general results~\cite{BPRV}.  Consider a
dynamical system $ {\bf X}(0) \to {\bf X}(t)=U^t {\bf X}(0)$ whose
time evolution can also be not completely deterministic (\textit{e.g.}
stochastic differential equations), with states ${\bf X}$ belonging to
a $N$-dimensional vector space.  We assume a) the existence of an
invariant probability distribution $\rho({\bf X})$, for which an
``absolute continuity'' condition is required (see~\cite{BPRV} for details), and b) the mixing
character of the system (from which its ergodicity follows). In our
stochastic model the two above requests hold. Under these hypotheses,
it is possible to derive (for details see~\cite{DH75,FIV90,BPRV}) the
following generalized FD relation, valid when considering the
perturbation at time $0$ of a coordinate $X_j$:
\begin{equation}
\label{eq:fdt}
R_{i,j}(t) = 
\frac{\overline{\delta X_i(t)}}{\delta X_j(0)}=- \Biggl \langle X_i(t) \left.
 \frac{\partial \ln \rho({\bf X})} {\partial X_j} \right|_{t=0}
\Biggr  \rangle \, .
\end{equation}
In the case of thermostatted Hamiltonian systems, on the other side,
one has that $\rho({\bf q},{\bf  p}) \propto \exp(-\beta
\mathcal{H}({\bf q},{\bf p}))$. 
From formula~(\ref{eq:fdt}),therefore, one has that 
\begin{equation} \label{eq:einstein}
R_{V,V}=\frac{ \langle V(t) V(0) \rangle}{\langle V(0)^2 \rangle}.
\end{equation}
With a slight abuse of terminology, we will use the form ``Einstein
relation'' to denote the time dependent Eq.~(\ref{eq:einstein}). Let
us note that its validity is a consequence of the Gaussian statistics
of the velocity and the factorization of the stationary probability
distribution, i.e. positions and velocities are independent.  In
non-Hamiltonian systems, the shape of
$\rho({\bf x})$ is not known in general, therefore~(\ref{eq:fdt}) does not give a
straightforward information. Nevertheless it can be exploited to get
an interpretation of the results of a linear response experiment. We
will analyze the response to small perturbations in the stationary
state of a one-dimensional granular gas, discussing the response
properties of the stationary state with its many ``anomalies'' with
respect to an equilibrium state.

We stress that the regimes considered
here are always ergodic: this is a relevant difference with respect to
the studies on the violations of the Fluctuation-Response
relation, which considered glassy systems in the non-ergodic (aging)
phase~\cite{bckm98}.

\paragraph{The model}

The model considered here consists of a gas of $N$ inelastic hard rods of mass $1$,
of linear size $d$, moving on a ring of length $L$. The rods interact
also with a heating bath which mimics the effect of an irregular
vibration injecting energy in the system. Until a collision occurs,
the position $x_i$ and the  velocity $v_i$ of $i$-th rod obeys the following equations:
\begin{equation}
\frac{dx_i(t)}{dt}=v_i(t),\;\;\;\;\;\;
\frac{dv_i(t)}{dt}=-\frac{v_i}{\tau_b}+\sqrt{\frac{2T_b}{\tau_b}}\eta_i(t),
\end{equation}
where $\eta_i(t)$ is a Gaussian white noise with $\langle \eta_i(t)
\rangle=0$ and $\langle \eta_i(t)
\eta_j(t')  \rangle=\delta_{ij}\delta(t-t')$. When two rods $i$ and $j$ come
into contact, their velocities $v_i$ and $v_j$ are instantaneously changed into $v_i'$ and $v_j'$ with the
following rule:
\begin{equation}
v_i'=v_i-\frac{1+r}{2}(v_i-v_j),\;\;\;\;\;\;\;
v_j'=v_j+\frac{1+r}{2}(v_i-v_j).
\end{equation}
The meanings of $\tau_b$ and $T_b$ are those of a typical
thermalization time and a temperature, respectively, obtained if the
system is elastic ($r=1$). The coefficient of restitution $r \in
[0,1]$ determines the degree of inelasticity: after a collision, a
fraction proportional to $1-r^2$ of the relative kinetic energy
(i.e. kinetic energy in the center of mass frame) of the two particles
is lost. When the particles are homogeneously distributed along the
ring, the mean free path is given by $\lambda=1/n-d=(1-\phi)/n$ where
$n=N/L$ is the number density and $\phi=nd$ is the
occupied volume fraction. The mean free time $\tau_c$ is roughly estimated as
$\lambda/\sqrt{T_g}$. In the rest of the paper we will tune $n$ or
$\tau_b$, keeping fixed $T_b=1$, in order to change the ratio between
characteristic times $\alpha=\tau_c/\tau_b$. For any value of $r$ or
$\alpha$, the system reaches a statistically stationary regime where a
kinetic temperature, denoted as ``granular temperature'', $T_g=\langle
v^2 \rangle$ can be measured.  When $\alpha \gg 1$, the coupling with
the thermostat dominates the dynamics of the rods: they therefore
remain thermalized and the system results at equilibrium at
temperature $T_b$: only spatial (rod-rod) correlations are expected at
equilibrium, while velocities are not correlated, i.e. the global
phase space probability distribution function (pdf) factorizes as:
\begin{equation} \label{fact}
\rho(\{x_i\},\{v_i\})=\rho_x(\{x_i\})\prod_{i=1}^Np_v(v_i),
\end{equation}
with $p_v(v)$ a Gaussian distribution with variance $T_b$.  On the
contrary, when $\alpha \ll 1$, the effect of inelastic collision is
strong enough to draw the system in a non-equilibrium stationary state
whose properties are known from previous studies~\cite{WM96,PLMPV98,PLMV99,NE98,NETP99}.
Non-Gaussian single particle velocity distributions and correlations
among  velocities and positions are the most relevant; these
anomalies with respect to equilibrium become more and more pronounced
as $\alpha$ or $r$ are reduced. As a matter of fact, in this regime it
is not correct to assume a factorization of the kind of
Eq.~(\ref{fact}), and the single particle velocity distribution, which
is non-Gaussian, represents only a projection on a single degree of
freedom of the full phase-space measure. We will see that the
non-gaussianity of velocities is by far less important that the lack
of factorization, which becomes relevant when the system is not dilute
enough and which makes the Einstein relation~(\ref{eq:einstein}) fail.


\begin{figure}[htbp]
\begin{center}
\includegraphics[width=9cm,clip=true]{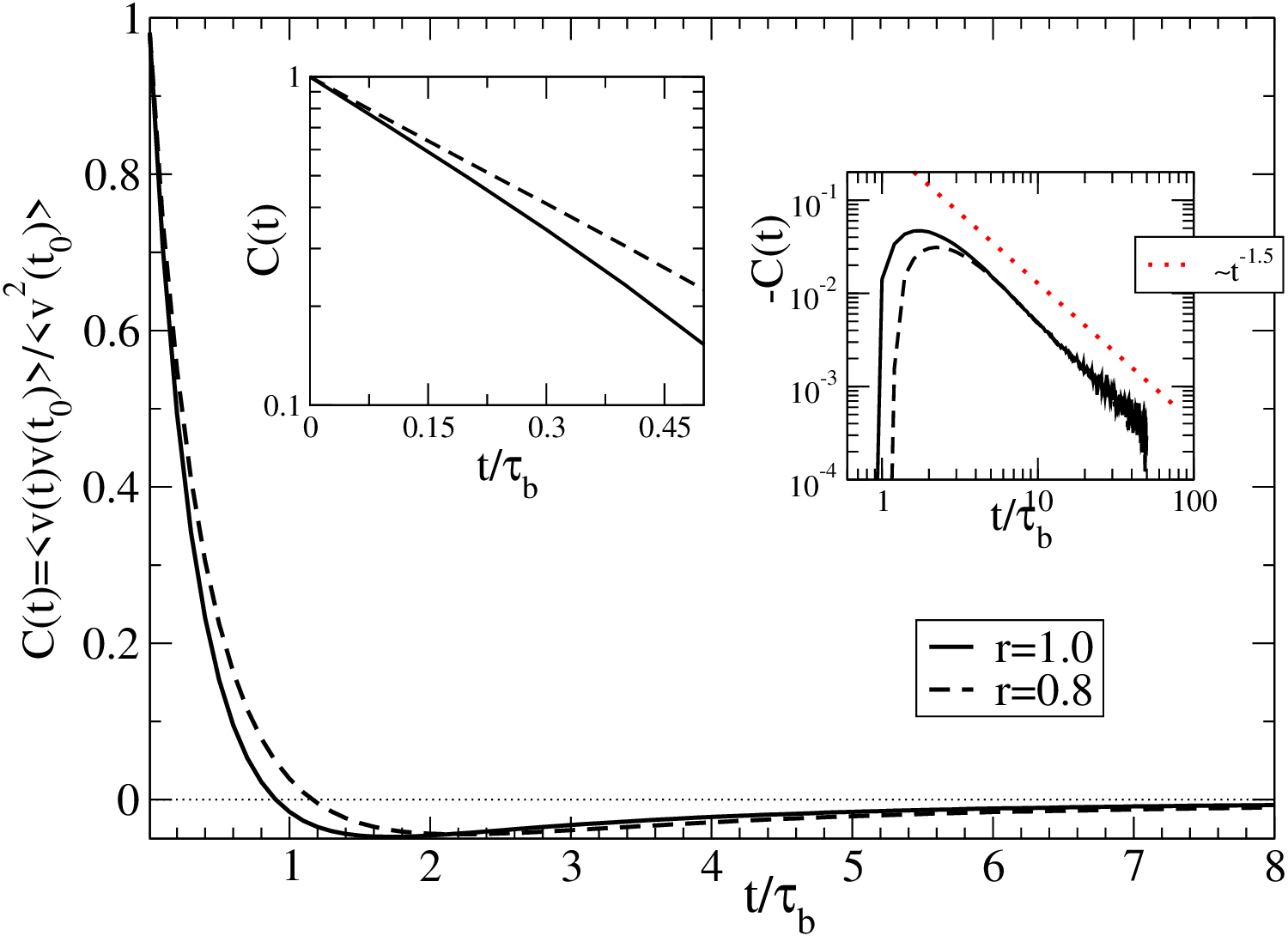}
\caption{Plot of the normalized autocorrelation $C(t)$ versus time,
  for two cases, one elastic (full line) and the other inelastic (dashed line). In the left
  inset we show a blow-up of the exponential decay at early times. In
  the right inset you can find a blow up in log-log scale of the
  negative tail, together with a power law decay $t^{-3/2}$. Here
  $\phi=0.1$, and $\alpha \approx 0.9$. \label{fig:1}}
\end{center}
\end{figure}

\paragraph{The velocity autocorrelation function}

In Figure~\ref{fig:1} we show the normalized autocorrelation function:
$C(t)=C_v(t)/C_v(0)=\langle v(t)v(0) \rangle/T_g$ for the velocity of
a tagged particle (a tracer with the same properties of other
particles). In both elastic and inelastic experiments, $C(t)$ presents
three main features: a) an exponential decay at early times, b) a
negative minimum and c) asymptotically a power-law decay $C(t) \sim
-t^{-3/2}$. The negative minimum is necessary to have subdiffusion,
i.e. $D=\int_0^\infty C(t)=0$, while the final power-law decay with
$3/2$ exponent is necessary to have $\langle x^2(t) \rangle \sim
t^{1/2}$. The initial exponential decay $C(t) \sim \exp(-t/t_{corr})$
has a more subtle nature. \forget{: it is typical of dilute systems
when dimensionlity is greater than $1$.} In $1D$ one can argue that
the tracer ``discovers'' the geometrical contraint after a long time.
\forget{,so that dimensionality is irrelevant for the initial decay of
$C(t)$.} However, calculations based on collisions between
non-correlated particles lead to wrong predictions for
$t_{corr}$. Since this point is not closely related to the FD
relation, we do not discuss it in detail. Here we do not show the mean
squared displacement as a function of time, already detailed
in~\cite{CDMP04}: however the single-file diffusion scenario $\langle
x^2(t) \rangle \sim t^{1/2}$ holds for any value of $r$, $\alpha$ and
$\phi$.

\paragraph{The response to an impulsive perturbation}

\begin{figure}[hbp]
\begin{center}
\includegraphics[width=7cm,clip=true]{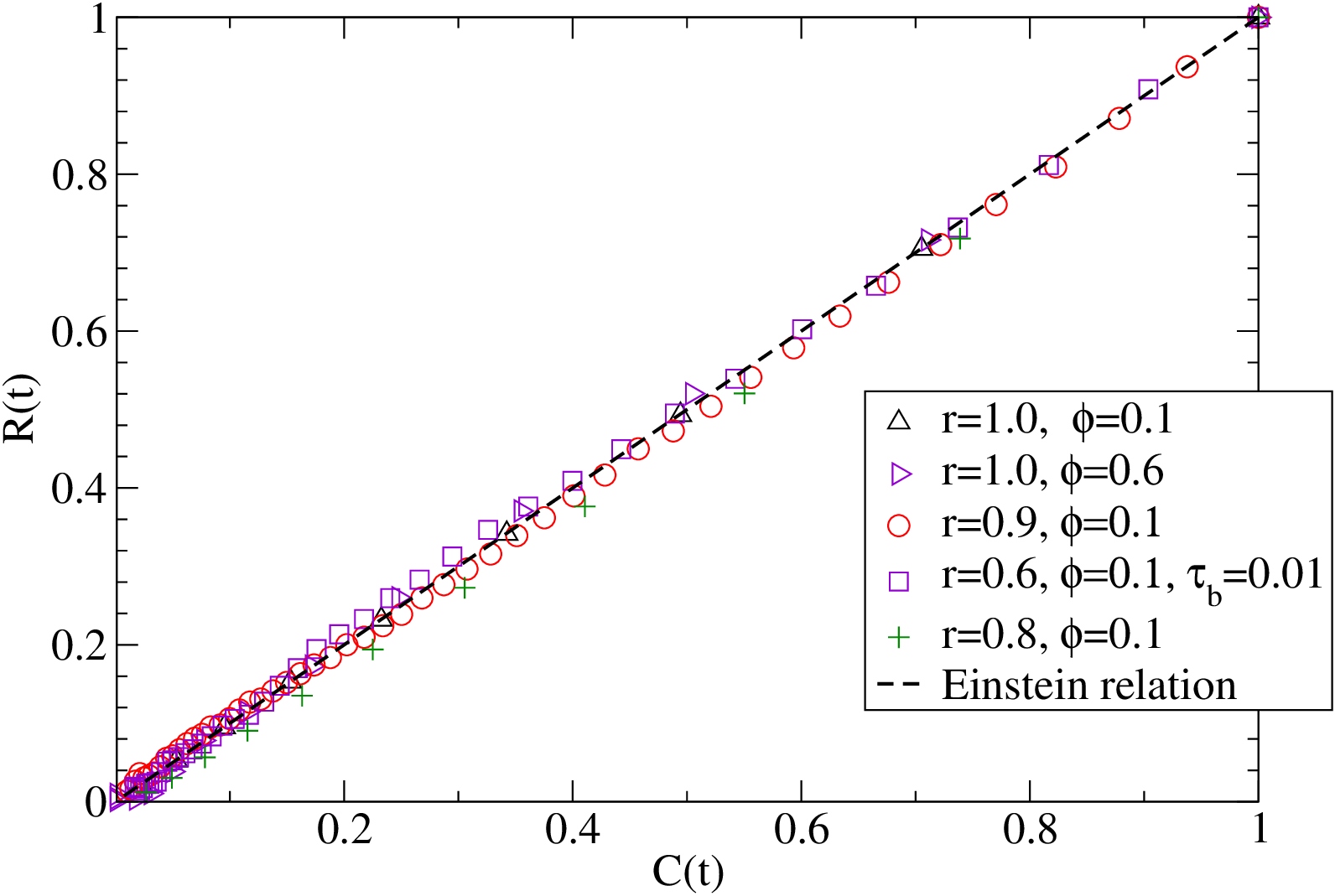}
\includegraphics[width=6.7cm,clip=true]{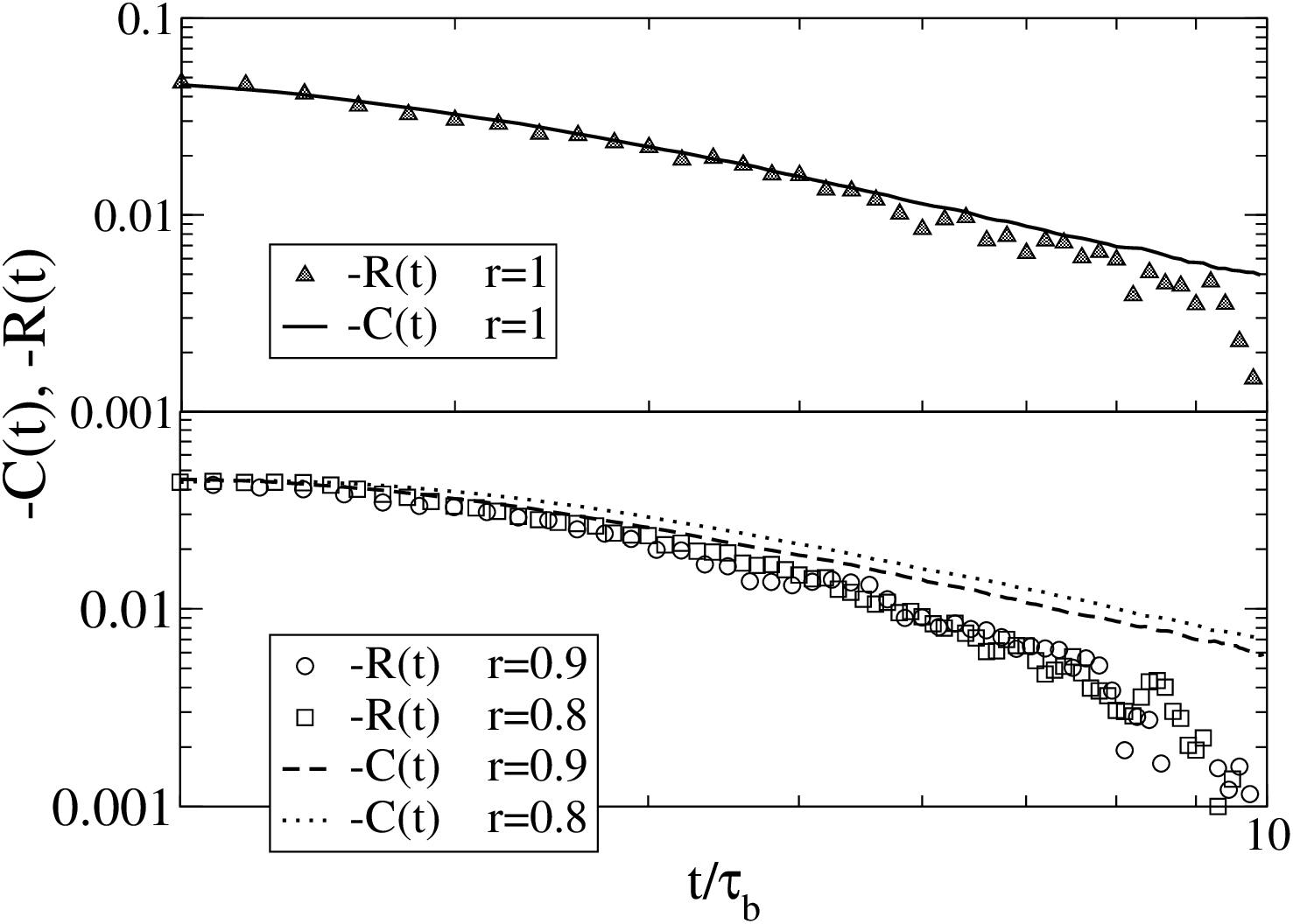}
\caption{Left: parametric plot of response $R(t)$ versus normalized
autocorrelation $C(t)$. The dashed line marks the Einstein relation $R
\equiv C$. Where not specified, $\tau_b=1$. Right: $-C(t)$ and $-R(t)$
versus $t$ for elastic and inelastic cases at late times, with
$\phi=0.1$ and $\tau_b=1$.  \label{fig:2}}
\end{center}
\end{figure}

The response to an impulsive perturbation is shown in
Figure~\ref{fig:2} for some choices of parameters. We have used a
standard recipe to have a clean measure of response~\cite{CJM79}: the
system is let thermalize, then at time $t_0$ is cloned. The original
system evolves without perturbation, the copy is perturbed, i.e. the
tagged tracer receives a small kick $v(t_0) \to v'(t_0)=v(t_0)+\delta
v$ with $\delta v \ll \sqrt{T_g}$ to ensure linearity of the
response. Then the copy is evolved using the same noise realization as
for the original system and the response is given by the dynamical
average $R(t)=\overline{(v'(t_0+t)-v(t_0+t))}/\delta v$ over many
realizations of the experiment. In Figure~\ref{fig:2} we show
representative cases where the Einstein relation $R(t)=C(t)$ is
verified within numerical precision. This happens for elastic cases,
or cases at low inelasticity $1-r \ll 1$ and low packing fraction, and
also for cases at high inelasticity, provided that $\tau_b \ll
\tau_c$. This last setup corresponds to a very fast action of the
thermal bath which practically removes the effects of inelastic
collisions. Similar results have been obtained, previously, for $2d$
driven granular gases~\cite{PBL02,BLP04,G04,BBDLMP05,PBV07}. As shown
in the top-right frame, for the elastic case, the relation $R(t)=C(t)$
is fairly verified also at late times in the power-law tail. The
inelastic case (see bottom-right frame) displays a small violation at
such large times: note that this small violation corresponds to both
$R(t)$ and $C(t)$ very close to zero.

\begin{figure}[htbp]
\begin{center}
\includegraphics[width=6.7cm,clip=true]{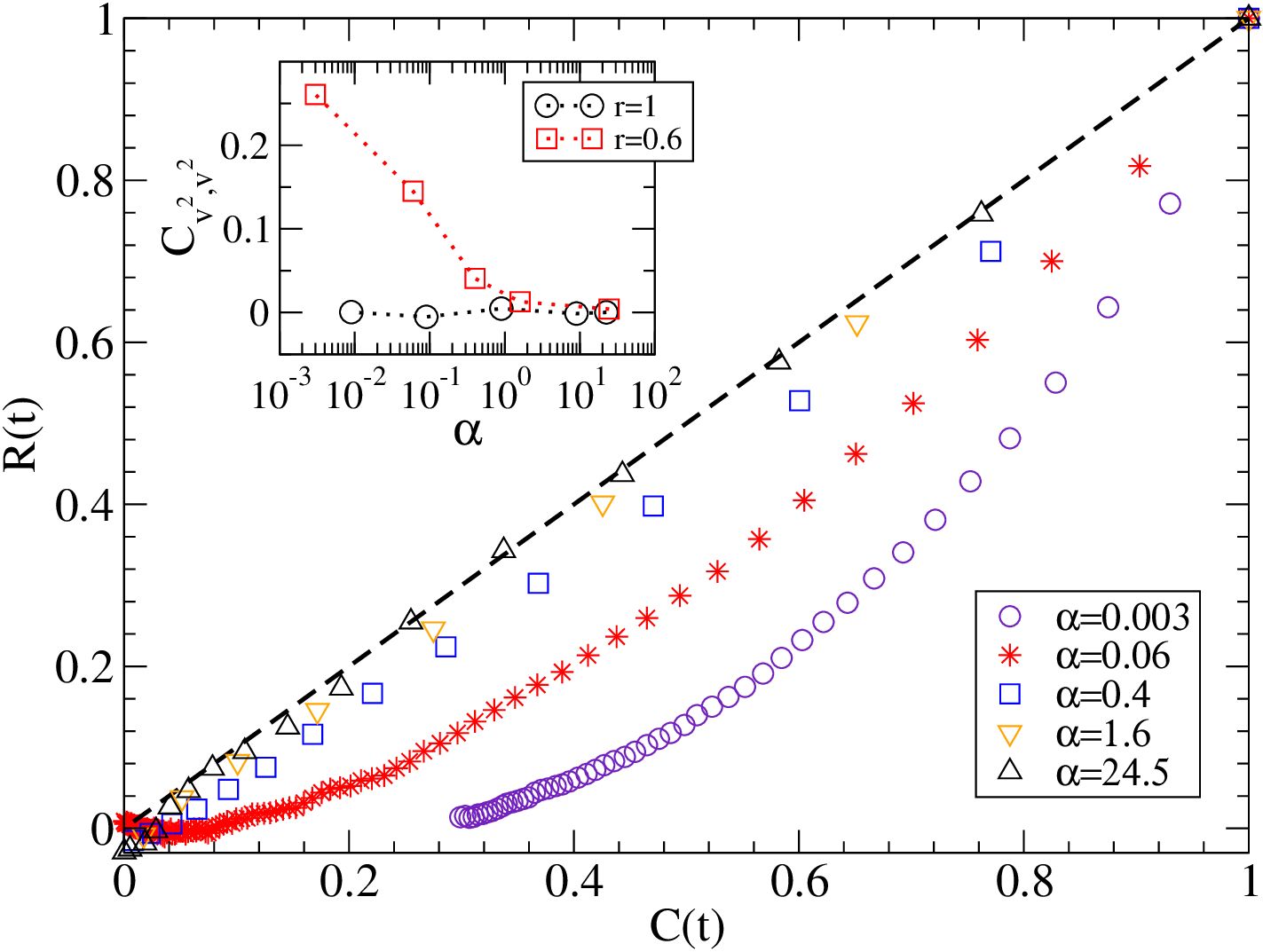}
\includegraphics[width=6.7cm,clip=true]{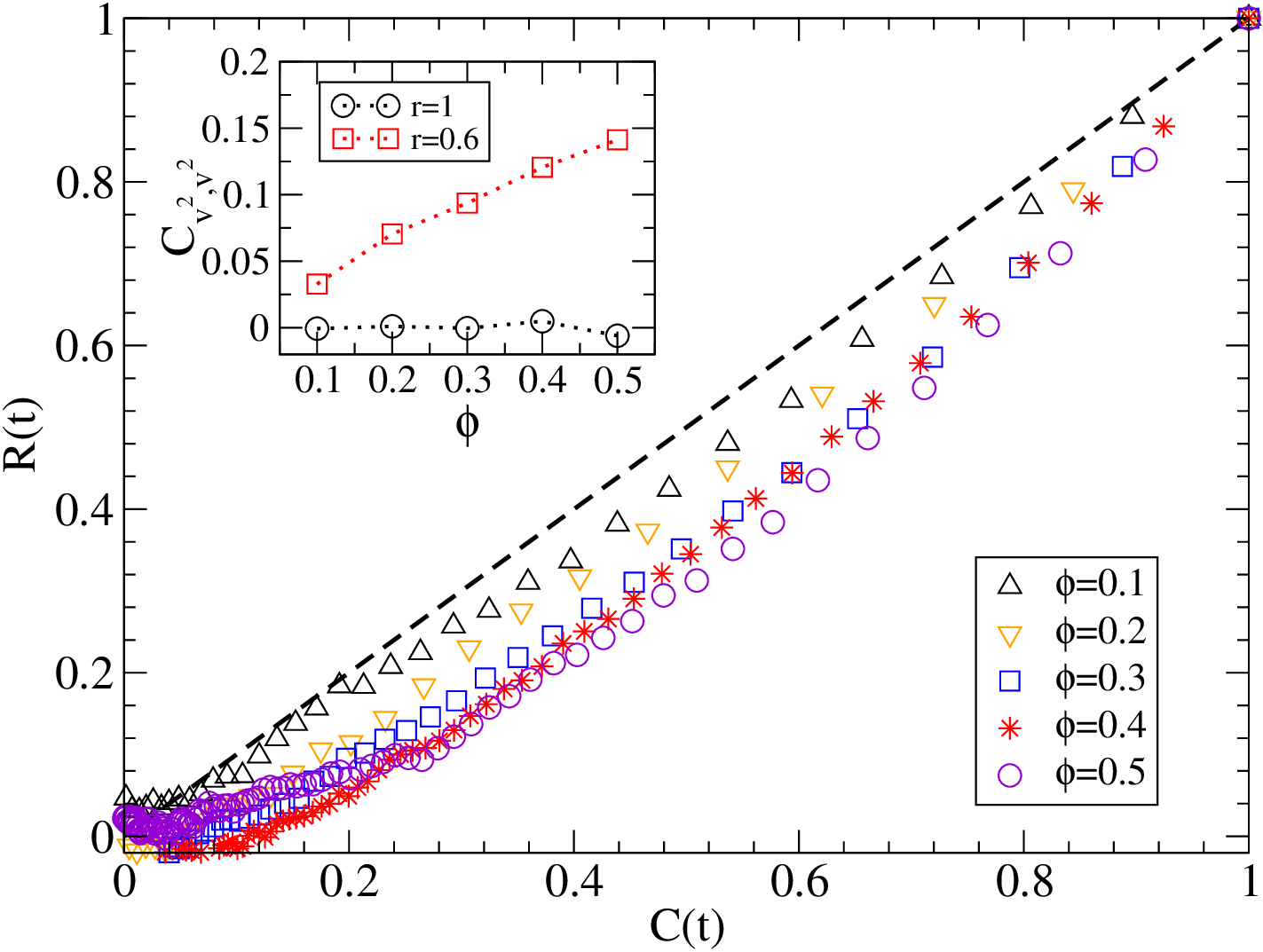}
\caption{Parametric plot of response $R(t)$ versus normalized
autocorrelation $C(t)$. The dashed line is the Einstein relation $R
\equiv C$. All data are obtained with restitution coefficient
$r=0.6$. On the left: the packing fraction is constant $\phi=0.1$ and
$\tau_b$ is changed, resulting in different values of $\tau_c$. The
ratio $\alpha=\tau_c/\tau_b$ is given for simplicity\label{fig:3}. On
the right: $\tau_b=1$ is kept constant, while $\phi$ is changed. In the
insets the correlator $C_{v^2,v^2}$, discussed in the text, is
displayed as a function of the varying parameter, for elastic and
inelastic systems.}
\end{center}
\end{figure}

In Figure~\ref{fig:3} the parametric plot of response versus
correlation is displayed for cases where the Einstein relation is no
more verified. The departure from the equality $R(t)=C(t)$ can be
quite strong: it increases with the packing fraction $\phi$, the
inelasticity $1-r$ and the rescaled bath time
$\tau_b/\tau_c=1/\alpha$. In all cases we observe $R(t) < C(t)$. In
Fig.~\ref{fig:3} we have stressed the dependence on $\alpha$, which
can be tuned changing $\tau_b$ at fixed $r$ and $\phi$.  In all
experiments we have verified to be in the linear response regime.

\paragraph{Origin of the violation of the Einstein relation}

As anticipated in the description of the model, and in agreement with
the observation done in~\cite{PBV07}, the Einstein relation no more
holds when the factorization of the phase-space pdf expressed by
Eq.~(\ref{fact}) is violated. For reasons of space we do not show the
probability density function of one-particle velocities, which are not
far from the Maxwell-Boltzmann distribution. Violations of Gaussianity
have been shown in~\cite{PBV07} to be not relevant for the FD
relation, because autocorrelations at different orders are almost
proportional, i.e. $\langle v(0)v(t)\rangle/\langle v^2\rangle \approx
\langle v(0)^2 v(t) \rangle/\langle |v|^3 \rangle \approx \langle v(0)^3
v(t) \rangle/\langle v^4 \rangle$ etc.  This is confirmed by Direct
Monte Carlo simulations, where an almost perfect factorization of the
degrees of freedom in the phase-space pdf is satisfied: in such
simulations, even with a stronger departure from Gaussianity, the
Einstein relation always holds.

Many ways of characterizing the breakdown of phase-space factorization
can be employed. A simple one is displayed in the inset of
Fig.~\ref{fig:3}:
\begin{equation}
C_{v^2,v^2}=\frac{\langle \delta v_i^2 \delta v^2_{i+1}\rangle}{\langle \delta v_i^4 \rangle},
\end{equation}
where $\delta v_i^2=v_{i}^2-T_g$. When $C_{v^2,v^2}>0$, the squared
velocities of two adjacent particles are correlated. It is evident
that this correlation increases when $\alpha$ is decreased. The same is
observed tuning the other parameters, such as decreasing $r$ or
increasing $\phi$.

\paragraph{Conclusions}

Drawing the conclusions, we stress the twofold nature of this
study. On one side, for the elastic single-file diffusion, which is a
less abstract model than fractional Fokker-Planck, we have obtained a
good agreement between $R(t)$ and $C(t)$, in all time ranges,
confirming the validity of the FD (``Einstein'') relation. On the
other side we have explored the effects of inelasticity: in this case
one has a non-equilibrium stationary state where strong correlations
among different particles are present, therefore the
factorization~(\ref{fact}) fails and only a more general FD
relation~(\ref{eq:fdt}) holds. At small inelasticity, small packing
fraction and/or for fast thermostats, the Einstein relation is
recovered, because the lack of factorization is weak, as previously
observed in $2d$ granular
gases~\cite{PBL02,BLP04,SL04,G04,BBDLMP05,PBV07}. A quantitative
characterization of the departure from factorization is under
investigation, with the aim of proposing, as a first step, a joint
two-particles (first neighbours) velocity distribution: we expect to
obtain, from this study, a first explicit correction formula to the Einstein
relation.

 \forget{On the
contrary, in non-factorizable situations, the Einstein relation fails
and the response always appear to be smaller than the velocity
autocorrelation.}

\vspace{.5cm}

\bibliographystyle{unsrt}
\bibliography{fluct.bib}

\forget{

[BF98] E. Barkai and V. N. Fleurov
``Generalized Einstein relation: A stochastic modeling approach''
Phys. Rev. E {\bf 58}, 1296 (1998).\\

[TFWG94] G. Trefan, E. Floriani, B.J. West and P. Grigolini
``Dynamical approach to anomalous diffusion:
Response of Levy processes to a perturbation''
Phys. Rev. E {\bf 50}, 2564 (1994).\\

[GSGWS96] Q. Gu, E.A. Schiff, S. Grebner, F. Wang and R. Schwarz
``Non-Gaussian transport measurements and the Einstein relation
 in amorphous silicon''
Phys. Rev. Lett. {\bf 76}, 3196 (1996).\\

[L98] P. Leboeuf,
``Normal and anomalous diffusion in a deterministic area-preserving map''
Physica D {\bf 116}, 8 (1998).\\

[BCVV95] L. Biferale, A. Crisanti, M. Vergassola, and A. Vulpiani
``Eddy diffusivities in scalar transport''
Phys. Fluids {\bf 7}, 2725 (1995).\\
}

\end{document}